%% file: main.tex
\documentclass{IEEEcsmag}

\usepackage[colorlinks,urlcolor=blue,linkcolor=blue,citecolor=blue]{hyperref}

\usepackage{soul}
\usepackage{upmath}
 \usepackage{fancybox}
 \usepackage{lmodern}
\usepackage{tcolorbox}
\usepackage{verbatim}
\usepackage[normalem]{ulem}
\usepackage{xspace}
\usepackage{amsfonts}
\usepackage{amsmath}

\jvol{XX}
\jnum{XX}
\paper{8}
\jmonth{May/June}
\jname{IT Professional}
\pubyear{2022}

\newcommand{\revise}[1]{#1}

\newcommand{\change}[1]{#1}

\definecolor{alizarin}{rgb}{0.82, 0.1, 0.26}
\definecolor{junglegreen}{rgb}{0.16, 0.67, 0.53}
\newcommand{\systemname}{\texttt{SUBPLEX}\xspace}
\begin{document}

\input{macros}

\sptitle{Department: Head}
\editor{Editor: Name, xxxx@email}

\title{\systemname: A Visual Analytics Approach to Understand Local Model Explanations at the Subpopulation Level}

\author{Jun Yuan}
\affil{New York University}

\author{Gromit Yeuk-Yin Chan}
\affil{New York University}

\author{Brian Barr}
\affil{Capital One}

\author{Kyle Overton}
\affil{Capital One}

\author{Kim Rees}
\affil{Capital One}

\author{{L}uis Gustavo Nonato}
\affil{Universidade de S\~ao Paulo}

\author{Enrico Bertini}
\affil{Northeastern University}

\author{Claudio T. Silva}
\affil{New York University}

\markboth{Department Head}{Paper title}

\begin{abstract}
\input{section/00-abstract}
\end{abstract}

\maketitle

\input{section/01-intro}

\input{section/02-background}
\input{section/04-method}
\input{section/03-pipeline}
\input{section/05-vis}
\input{section/06-eval}
\input{section/07-discuss}
\input{section/08-limit+conclude}

\bibliographystyle{IEEEtran}
\bibliography{reference}

\begin{IEEEbiography}{Jun Yuan}{\,} is currently a Ph.D. candidate in the Department of Computer Science and Engineering at New York University. Her research interest lies in the intersection of Data Visualization and Explainable AI (xAI). Contact her at junyuan@nyu.edu.
\end{IEEEbiography}

\begin{IEEEbiography}{Gromit Yeuk-Yin Chan}{\,} received his Ph.D. degree in Computer Science from New York University. He is broadly interested in the machine learning aspect of interactive visualization. Contact him at gromit.chan@nyu.edu.
\end{IEEEbiography}

\begin{IEEEbiography}{Brian Bar}{\,} is an explainable AI researcher at Capital One. He received a Ph.D. degree in Mechanical Engineering from Clarkson University. His research interests include deep learning on tabular data, and local attribution methods. Contact him at brian.barr@capitalone.com.
\end{IEEEbiography}

\begin{IEEEbiography}{Kyle Overton}{\,} is a Visual/UX designer at Capital One. He received a M.S. degree in Human-Computer Interaction from Indiana University Bloomington. Contact him at kyle.overton@capitalone.com.
\end{IEEEbiography}

\begin{IEEEbiography}{Kim Rees}{\,} is currently the head of Data Experience Design at Capital One. She established the Responsible AI program and leads the Data Privacy design team in Capital One. Contact her at kim.rees@capitalone.com.
\end{IEEEbiography}

\begin{IEEEbiography}{Luis Gustavo Nonato}{\,} is a full professor at Universidade de S\~ao Paulo. His research interest includes visualization, visual analytics, machine learning, data science, and geometric computing. Contact him at gnonato@icmc.usp.br.
\end{IEEEbiography}

\begin{IEEEbiography}{Enrico Bertini}{\,} is an Associate Professor at Northeastern University. His research focuses on the development and evaluation of interactive visual interfaces to help people reason with data and machine learning models. Contact him at e.bertini@northeastern.edu.
\end{IEEEbiography}

\begin{IEEEbiography}{Claudio T. Silva}{\,} is Institute Professor of Computer Science and Engineering at NYU Tandon School of Engineering, and Professor of Data Science at the NYU Center for Data Science. His research interest includes visualization and graphics, geometric computing, data science, urban computing, sports analytics. Contact him at csilva@nyu.edu.
\end{IEEEbiography}

\end{document}

%% file: macros.tex
\renewcommand{\paragraph}[1]{\vspace{0pt} #1}
\newcommand{\paragraphem}[1]{\vspace{0pt}\noindent \textbf{#1}}
\newcommand{\casestudy}[2]{\vspace{0.1cm}\noindent \textbf{#1}\\ \vspace{0cm}\noindent #2}
\newcommand{\eg}{e.g.,\xspace}
\newcommand{\ie}{i.e.,\xspace}
\newcommand{\etc}{etc.\xspace}
\newcommand{\hidecomment}[1]{}
\newcommand{\savespace}[1]{}
\newcommand{\denselist}{\itemsep 0pt\parsep=1pt\partopsep 0pt}
\newcommand{\decsp}{\vspace{-0.1in}}
\newcommand{\itemspace}{\vspace{-.15cm}}

\newcommand{\highlight}[1]{#1}

\newcommand{\header}[1]{\multicolumn{1}{c}{\textbf{\Shortunderstack{#1}}}}
\newcommand{\headerfirst}[1]{\multicolumn{1}{|c}{\textbf{\Shortunderstack{#1}}}}
\newcommand{\headerlast}[1]{\multicolumn{1}{c|}{\textbf{\Shortunderstack{#1}}}}
\newcommand{\headergap}[1]{\multicolumn{1}{c}{\addstackgap{\textbf{\Shortunderstack{#1}}}}}

\newcommand{\etal}{\textit{et al.}\xspace}

\newcommand{\Dspace}{$\mathbb D$\xspace}
\newcommand{\Sspace}{\mathbb S}
\newcommand{\Tspace}{\mathbb T}
\newcommand{\Rspace}{\mathbb R}

\newcommand{\task}[1]{\textbf{T.#1}}
\newcommand{\tasks}[1]{Tasks~#1\xspace}

\newcommand\crule[3][black]{\textcolor{#1}{\rule{#2}{#3}}}

\newcommand{\clabel}[1]{\textcircled{\tiny{\textbf{#1}}}}
\newcommand{\cblabel}[1]{\textcircled{\small{#1}}}

\newcommand*\circled[2]{\tikz[baseline=(char.base)]{
            \node[shape=circle,draw,inner sep=0.5pt,fill={#2},scale=0.75] (char) {\tiny{#1}};}}
\newcommand{\colorrgb}[3]{rgb,255:red,#1; green,#2; blue,#3}

\newcommand{\todo}[1][red]{\textcolor{#1}}

\definecolor{lavenderblush}{rgb}{1.0, 0.94, 0.96}
\definecolor{alizarin}{rgb}{0.82, 0.1, 0.26}

\newtcbox{\boxcolor}[1][lavenderblush]{on line,
arc=3pt,colback=#1,colframe=white,
before upper={\rule[-3pt]{0pt}{10pt}},boxrule=1pt,
boxsep=0pt,left=2pt,right=2pt,top=1pt,bottom=.5pt}

\newcommand{\boxtext}[2]{\boxcolor[#1]{\small \textsf{\textsc{#2}}}}

\newcommand{\ra}[1]{\renewcommand{\arraystretch}{#1}}

\newcommand{\punchline}[1]{}

%% file: section/00-abstract.tex
Understanding the interpretation of machine learning (ML) models has been of paramount importance when making decisions with societal impacts such as transport control, financial activities, and medical diagnosis. While local explanation techniques are popular methods to interpret ML models on a single instance, they do not scale to the understanding of a model's behavior on the whole dataset. 
In this paper, we outline the challenges and needs of visually analyzing local explanations and propose \texttt{SUBPLEX}, a visual analytics approach to help users understand local explanations with subpopulation visual analysis. \texttt{SUBPLEX} provides steerable clustering and projection visualization techniques that allow users to derive interpretable subpopulations of local explanations with users expertise. We evaluate our approach through two use cases and experts' feedback.


%% file: section/01-intro.tex
\punchline{Local attribution is popular for XAI. (LIME, DeepLIFT, Shap).}
\chapterinitial{Machine Learning (ML)} produces accurate prediction models that can be applied to address important societal problems.
While the advance of ML attracts a broader application in different domains, its increasing influence also raises a greater demand for transparency and interpretation of the decisions an ML model makes.
Thus, the machine learning community has devoted efforts to explaining an ML model's behavior, regarding or regardless of a model's internal structure. A popular approach is to generate a \textit{local explanation} per input to a model~\cite{kim2017interpretability,ribeiro2016should}, where the explanation is a feature vector that explains the importance of each feature on an input to the model's decision. 
These methods are popular due to accurate results and easy-to-understand representations for end users.

However, one challenge of using local explanation is the lack of a \textit{global view} on an ML model. By focusing on explaining a single instance out of all instances in the whole dataset, users might lose context on how the model behaves on a much larger population. 
Therefore, to understand a large number of local explanations, it is worthwhile to apply visual analysis~\cite{shneiderman2003eyes} since users can acquire a big picture of the ML model by exploring all local explanations of a dataset at once through their visual representations.
To be specific, users can visually group similar local explanations to gain an overview of different ways the model uses to make decisions. Then, they can understand each way of explanation by exploring each group of local explanations.
We call such a process ``visual analysis on local explanation sub-populations''.

\punchline{Sub-population visual analysis = projection and clustering, but simply applying them will not work well.}
To conduct such visual analysis, users can apply high dimensional data visualization techniques with well-studied methods such as clustering and projections~\cite{wenskovitch2017towards}. However, we find out that a straightforward solution of using clustering and projection on local explanations does not work well for the following reasons. 
First, as we will demonstrate in following sections, the data characteristics of local explanations often hinder the effectiveness of clustering and projection results from revealing an accurate overview of a model's behavior.
Second, through working with domain experts, we discovered that users need a human-in-the-loop mechanism to steer the local explanation explorations and identify important features among all the features generated in a local explanation. 
Third, users need the system to be embedded in their data science workflow.

\punchline{We provide a system to steer projection and clustering on local attributions.}
Having these challenges in mind, we designed \textsc{SUBPLEX},
a visual analytics approach that visualizes ML model local explanations at a subpopulation level.
It is designed through an iterative discussion with a team of data scientists.
We combine the concepts of subpopulation analysis in visualization and real industrial tasks on model interpretability for model developers to induce a feedback-driven analysis of subpopulation analysis on local explanations at scale.
In short, our contributions include:
\begin{enumerate}
    \item A human-in-the-loop framework for local explanation subpopulation analysis. 
    \item An interactive visual analytics tool embedded in Jupyter notebook, {\systemname}, based on an iterative design process.
    \item Two use cases and user feedback collection as evaluation of {\systemname}.
\end{enumerate}

%% file: section/02-background.tex
\section{Background and Related Work}
\label{sec:related}

\subsection{Local Explanations for Model Understanding}
To deal with the inherent complexity of ML models and their lack of transparency, a number of ``post-hoc'' explanation or ``attribution''  methods have been developed. 
In this work, we focus on the local explanations, which help humans understand how individual prediction is made. 

Current local explanations usually express the \textit{explanation} for a data point as a vector where each value inside the vector is a human-understandable object. For example, additive feature explanation methods like LIME \cite{ribeiro2016should}, SHAP \cite{lundberg2017unified}, and GAM \cite{hastie2017generalized} output the explanations as a list of feature importance.
Therefore, we can define an explanation $a$ for an input as a real valued explanation vector mapped to a feature space with $m$ features:
\begin{equation} \label{eqn:attribution}
    a = \left \{ w_1, w_2,...,w_m \right \} \forall w_i \in \mathbb{R}
\end{equation} 
where each weight $w_i$ represents the importance of a feature $W_i$.


In this work, we focus exclusively on local explanations in the format of the attribution vector $a$. We aim to help data scientists to discover and interpret subpopulations with different local explanation patterns.

\subsection{Visual Analytics of Local Explanations}
To interpret local explanations on a set of instances, people calculate aggregated explanation values, for example, the average of the absolute SHAP values ~\cite{molnar2019}. However, average values over the whole dataset can be misleading because a feature might only be ``important'' for a subgroup but not the whole dataset. 
Lundberg et al.~\cite{lundberg2018consistent} introduce the analysis of SHAP values in clusters. However, they only use a straightforward clustering approach. In our work, we support both automatic clustering and manual subpopulation creation. Recent visual analytics systems such as DECE~\cite{cheng2020dece} also helps understand model behavior on a subpopulation of data using rules. Our work, instead, focuses on local explanation vectors.

Recently, more visual analytics techniques have been proposed to help users analyze local explanations. Collaris \textit{et. al}~\cite{collaris2022comparative} introduced local/global contribution-value (LCV, GCV) plots by showing how a local explanation changes given different feature values. Similarly, ExplainExplore~\cite{collaris2020explainexplore} help users identify instances classified with similar reasons by plotting all the LIME~\cite{ribeiro2016should} values. Our approach, instead, automatically identifies subpopulations with similar local explanations and then visualizes aggregated local explanations at the subpopulation level.


Dimension reduction algorithms and clustering algorithms are both frequently used techniques in visual analytics for subpopulation-level data analysis. Recent surveys \cite{wenskovitch2017towards,nonato2018multidimensional} list how they are combined in practice. \change{In general, clustering implies visually highlighting similar instances (e.g., with the same color), while projections imply visualizing the samples as dots in a scatter plot with positions correlated to the similarities among the instances.} We apply the two methods in our work with steerable mechanisms focusing on the characteristics of local explanations to aid the analysis.

%% file: section/04-method.tex
\begin{figure*}
    \centering
    \includegraphics[width=\textwidth]{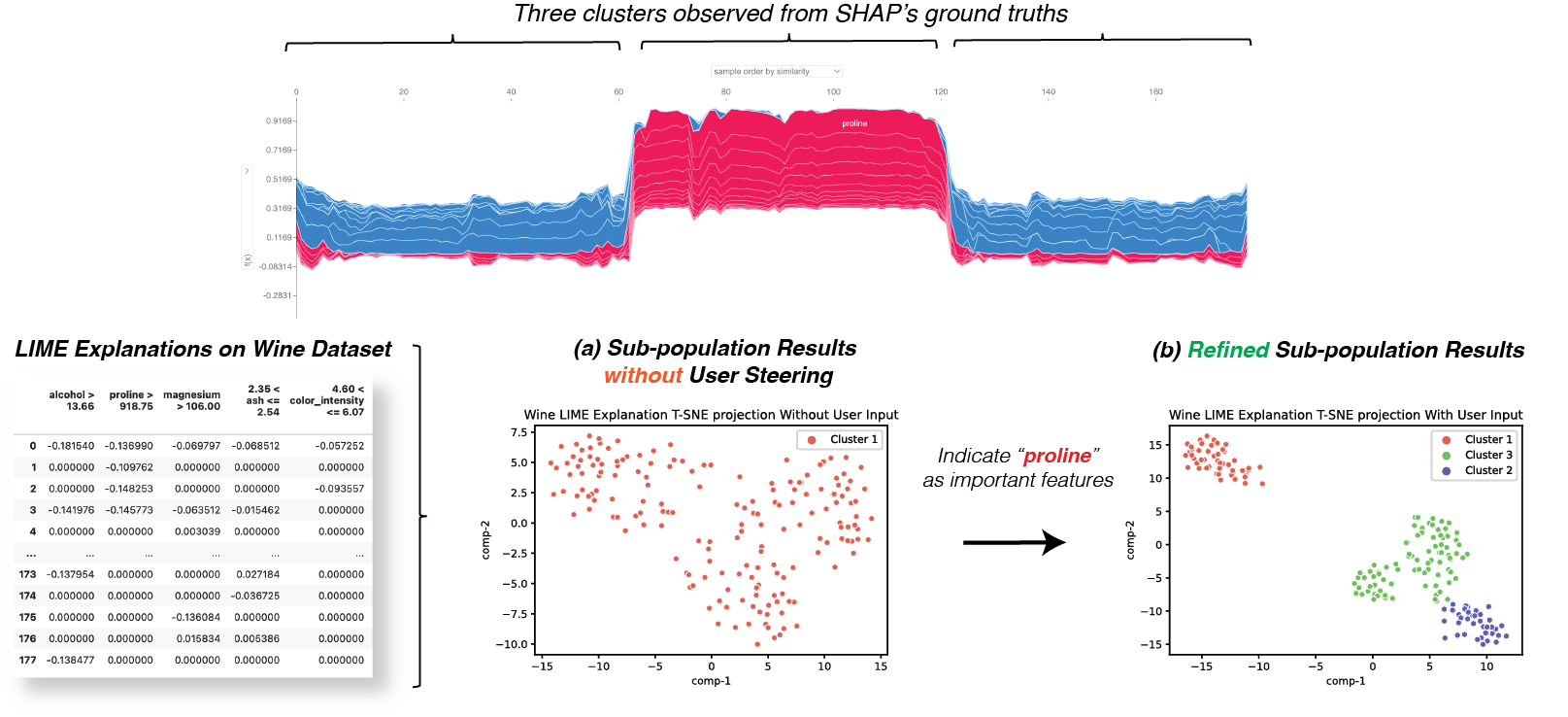}
    \caption{ 
    An illustration of refining LIME subpopulation results with a synthetic dataset. The ground truth of the classifier from SHAP contains \textit{three} decision boundaries. (a) The original subpopulation result of the explanations does not reflect the ground truth. (b) When selecting the features related to \textit{proline} which are the main rationale of the model, the subpopulation result reflects three groups and becomes consistent with the ground truth.
    }
    \label{fig:synthetic}
\end{figure*}

\section{Subpopulation Analysis on Local Explanations}
\label{sec:method}
In this section, we introduce the technical challenges of visually analyzing local explanations and propose a steerable subpopulation analysis pipeline to address the technical challenges.

\subsection{Clustering and Projections}
\punchline{Grouping = Similarity Among Explanations}
A subpopulation analysis implies that we are interested in grouping the local explanations across the dataset based on their similarities. Grouping allows us to aggregate the explanations and reduce the visual complexity to examine a large set of explanations simultaneously. In data visualization, the popular approaches are clustering and projections~\cite{wenskovitch2017towards} as we describe in the previous section.

While we might address our goal of analyzing local explanations at scale by directly applying these methods, our collaborations with domain experts revealed several technical challenges. These challenges required us to provide additional metrics and mechanisms to facilitate the visual analytics process. Before introducing the system design requirements in the next section, our technical challenges from the data (i.e., local explanations) perspective are as follows:
\begin{enumerate}
    \item \textit{ Local explanations are sparse and noisy.}
    The comparison between local explanations as a whole is not meaningful since some features are more important than others due to sparsity and noisiness in the explanation outputs. The distance metric for the subpopulation result should not treat features equally.
    \item \textit{Many features exist in the explanations to inspect or refine.} To identify important features to group explanations or discover the important features, users need to inspect many features. Without feature ranking or filtering, there will be lots of human effort in the analysis.
\end{enumerate}

To illustrate how the above shortcomings will affect the subpopulation results of local explanation, we present a result using a synthetic Wine classification dataset~\cite{Dua:2019} (Figure~\ref{fig:synthetic}). 
To obtain a ground truth on how many groups of rationale exist in the dataset and classifier, we train a random forest classifier and obtain a SHAP explanation~\cite{lundberg2017unified} on the data. We choose the random forest classifier as the illustration since the SHAP algorithm can generate accurate explanations and the number of subpopulation clusters on tree-based classifiers.
Upon knowing there are three clusters of explanations from SHAP's plot (Figure~\ref{fig:synthetic}), we generate other local explanations using LIME. The local explanations are harder to explore since they have 52 dimensions, and the values are sparse in general. We visualize the explanations with projections and colors generated from DBSCAN clustering algorithm to highlight the number of similar groups in the explanations. 
The results are shown in Figure~\ref{fig:synthetic}a. We cannot observe any significant group structure from either the clusters or projections due to the high dimensionality and sparsity from the LIME explanations. 

\subsection{Steering Subpopulation Results}
\punchline{Transform Similarity Matrix with Weights}
Seeing the above shortcomings of subpopulation analysis for local explanations, we are motivated to introduce human input to refine the results of the computed subpopulations. 
Noted that any pipeline of subpopulation techniques can be generalized by first constructing a distance measure among instances (i.e., local explanations) and then applying a subpopulation technique (e.g., MDS, tSNE, K-means clustering).
Thus, to generalize our approach for choosing projection and clustering techniques with flexibility, we propose a steerable technique that focuses on adjusting the distance measure with human input.

Since most of the clustering and projection techniques takes Euclidean distance between high dimensional vectors as inputs, we define a user-input-aware distance metric:
\begin{flalign}
    d(a_i,a_j) &= \nonumber\\ \sqrt{\sum_{k \in f_u}{\alpha({w_i}_k - {w_j}_k)^2}  + \sum_{k \in f_n}{\beta({w_i}_k - {w_j}_k)^2}}
\end{flalign}
where $f_u$ are the features in the explanations that the user thinks it is important, and $f_n$ are the remaining features. $\alpha$ and $\beta$ are the constants that we set up to adjust the influences of the features to reflect the user's emphasis. By default, $\alpha = 0.9$ and $\beta = 0.1$.   

To understand how it improves the result, we use the same synthetic dataset in Figure~\ref{fig:synthetic}. If we add the features that reflect the decision boundaries (i.e. ``proline'') to $f_u$, we can see the grouping becomes more related to the ground truths (Figure~\ref{fig:synthetic}b), where there exist three groups from the clustering and projection results.

\subsection{Recommending Features for Steering}
Since feature space might be huge, to assist users in refining the clustering results, we present feature suggestions and enable feature customization for cluster generation. 

We now define each feature across the whole explanation dataset as $W_i$, and our goal here is to provide a score for each feature (i.e., $score(W_i)$). 
When a clustering result is generated, we obtain a set of clusters $\mathcal{C} = \{C_1, C_2,...\}$, where $C$ represents a set of similar explanations.
Inspired by cluster evaluation methods~\cite{halkidi2002cluster}, we propose three properties to score features: 

\textbf{Property 1 (Consistency):} the variance of $W_i$ within each cluster. 
\begin{equation}
    cst(W_i) = \sum_{C\in\mathcal{C}}{\|C\|/var(\{{w_j}_i | a_j \in C\})}
\end{equation}
where $\|C\|$ is the number of instances in the cluster $C$. The higher the $cst(W_i)$, the more consistent the local explanations of feature $W_i$ are distributed in each cluster. 

\textbf{Property 2 (Separation):} the difference of values of $W_i$ across different clusters.
\begin{equation}
    sep(W_i) = var(\{mean(\{{w_j}_i | a_j \in C\})|C\in \mathcal{C} \})
\end{equation}
The higher the $sep(W_i)$ is, the more distinctive the local explanations of feature $W_i$ are in different clusters. 

\textbf{Property 3 (Importance):} the significance of $W_i$ in one or more clusters.
\begin{equation}
    impt(W_i) = max(mean(\{{w_j}_i | a_j \in C\}))
\end{equation}
The higher the $impt(W_i)$ is, the more important feature is in at least one cluster.

Under Properties 1-3, for a given feature $W_i$ and the present clusters, we calculate a feature score:
\begin{equation}
    score(W_i) = \overline{cst(W_i)} + \overline{sep(W_i)} + \overline{impt(W_i)}
\end{equation}
where $\overline{(\cdot)}$ normalizes the values within the range of 0 to 1.

Once the users run the clustering algorithm, we suggest the top k features with the highest feature scores (default $k=6$). 

%% file: section/03-pipeline.tex
\begin{figure*}
    \centering
    \includegraphics[width=\textwidth]{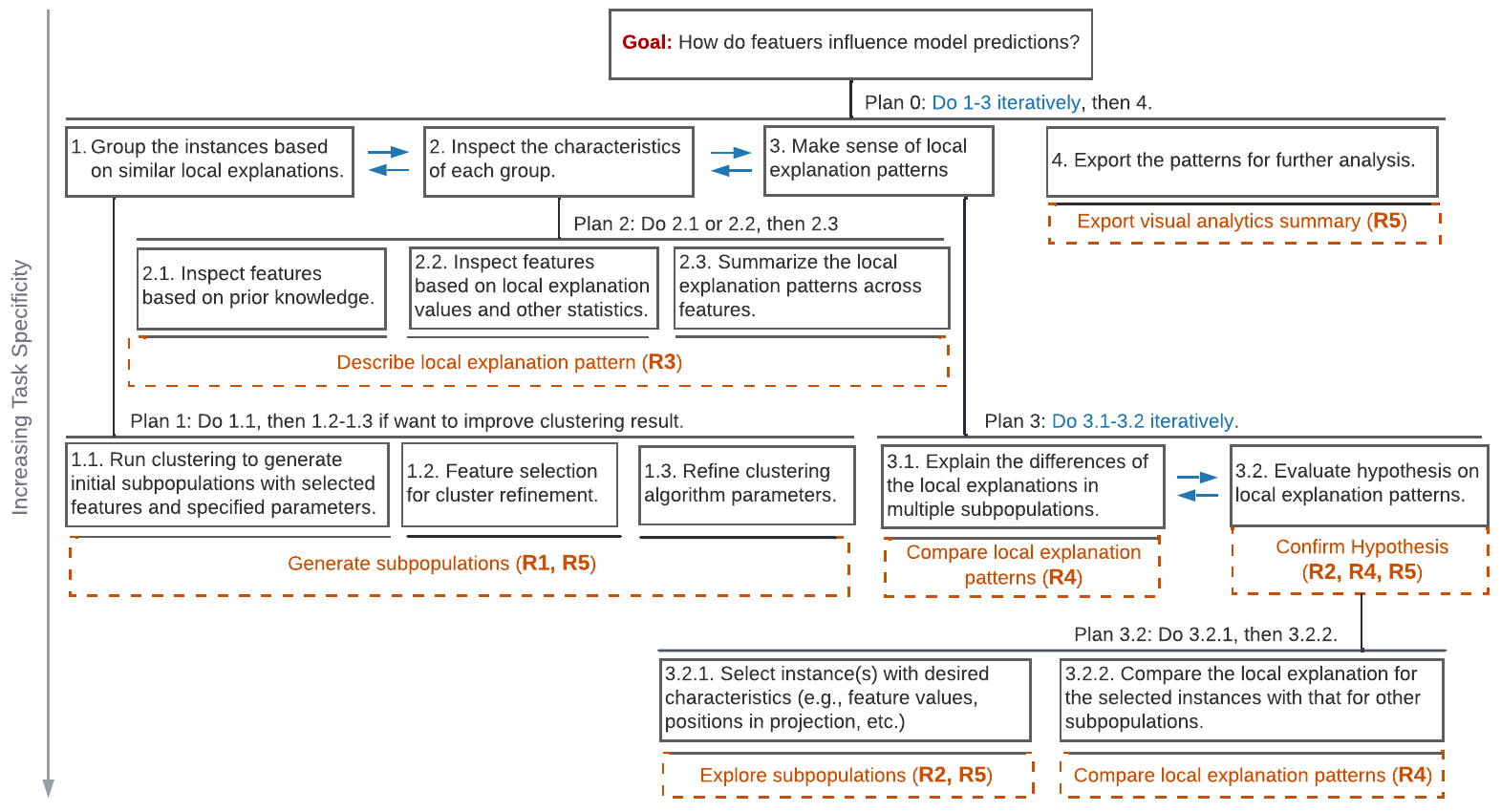}
    \caption{Hierarchical Task Abstraction (HTA) of local explanation analysis using box-and-line notation. We follow the standard conventions for hierarchical task analysis~\cite{kurniawan2004interaction} where tasks are represented by named boxes with a unique ID, which also indicates the hierarchical level of the task. Task abstraction based on ~\cite{lam2017bridging} are highlighted in orange. The related design requirements are denoted with each abstract task.}
    \label{fig:hta}
\end{figure*}

\section{Human-in-the-loop Pipeline}
\label{sec:pipeline}
In this section, we introduce the design rationale of {\systemname}. {\systemname} is developed as a part of an explainable artificial intelligence project. 
We held bi-weekly meetings with experts in machine learning and visualization for 12 months, at which a variety of design requirements were gathered. Together with the computation and steering mechanism in the previous section, we developed the visual analytics system via the iterative user-centered design process. Below we first connect the design goal with specific analytical tasks through hierarchical task analysis~\cite{salmon2010hierarchical} and task abstraction~\cite{lam2017bridging}. Then we list the design requirements and provide an overview of the resulting human-in-the-loop pipeline.

\subsection{Hierarchical Task Abstraction (HTA) and Design Requirements}

The goal of this work is to help data scientists understand model logic through local model explanations. More specifically, we want to help answer the question of ``How do different features influence the model predictions?''

To better support the goal of local explanation analysis, we conduct hierarchical task analysis (Figure~\ref{fig:hta})~\cite{salmon2010hierarchical,zhang2018idmvis} to understand the workflow. Then we apply the framework proposed by Lam et al.~\cite{lam2017bridging} to connect the analytical goal with abstract tasks to derive the design requirements. The four requirements (R1-4) from HTA with a general data science workflow requirement (R5) are as follows:



\textbf{R1: Guided feature selection for clustering refinement.}
To mitigate the effects of noisy features, we should guide users to identify features that lead to clear clusters.

\textbf{R2: Flexible support to analyze user-defined subpopulations.}
Data scientists may have instances they are interested in based on prior knowledge. Therefore, our tool should support flexible subpopulation creation by interactive selection and queries.

\textbf{R3: Visual assistance for identifying local explanation pattern in one subpopulation.}
Our tool should support interactive exploration to identify patterns, including the size of a cluster, feature distribution in a subpopulation, and instances in a subpopulation.

\textbf{R4: Visual comparison of local explanation patterns in different subpopulations.}
Our tool should support visual comparison of local explanation patterns for interpretation and validation of the local explanations.

\textbf{R5: Coding-environment-friendly visual analytics.} 
A general data science system requirement is to embed the system in computation notebooks to take the processed data for visualization and output results for later model development and deployment.

\subsection{System Overview }
\begin{figure*}
    \centering
    \includegraphics[width=\textwidth]{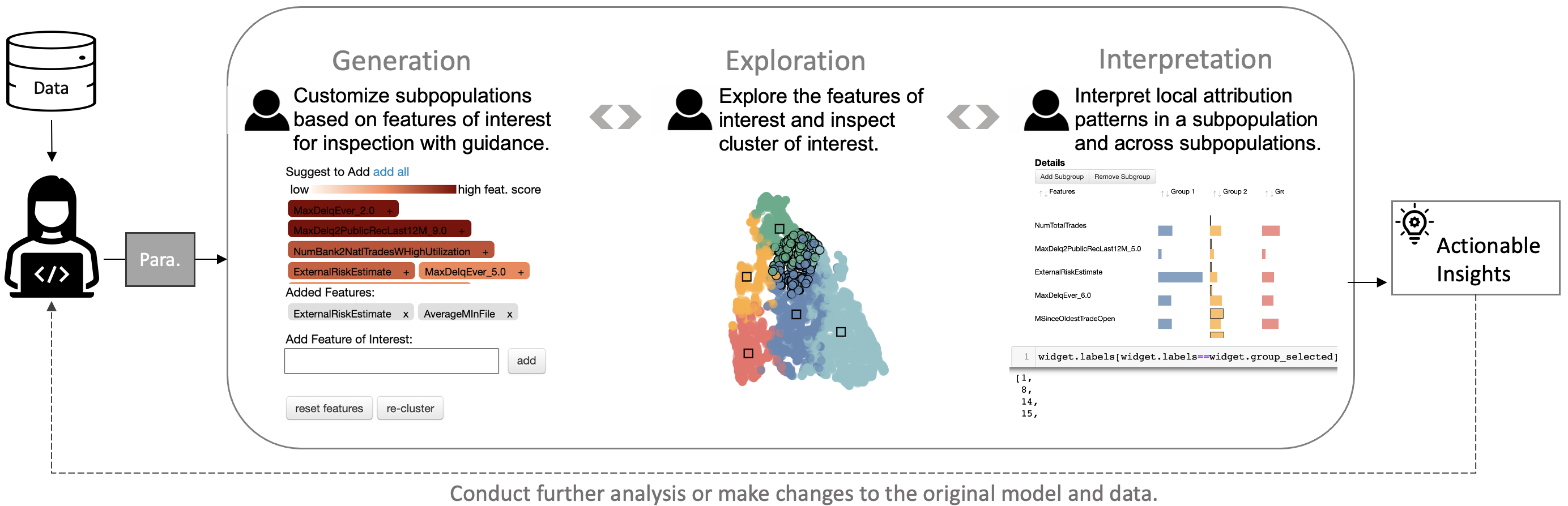}
    \caption{An overview of the human-in-the-loop pipeline for local explanation analysis in the coding environment.
    }
    \label{fig:pipeline}
\end{figure*}

Based on the HTA, we introduce a human-in-the-loop approach for subpopulation-level local explanation analysis. It contains three main stages as shown in Figure~\ref{fig:pipeline}. 

\textbf{Generation.} The first stage focuses on the automatic subpopulation generation (clustering) and refinement (R1). After the widget is initialized with the explanations to be analyzed, we visualize the clusters in the 2D space. We then suggest the top features with the highest feature scores.

\textbf{Exploration.} The second stage is to explore a specific subpopulation (R2, R3). Users can manually add/remove subpopulations (R2) by multiple means and identify the local explanation patterns (R3) in different subpopulations.

\textbf{Interpretation.} The third stage enables users to export intermediate results for further analysis in the notebook (R5) and test the hypothesis they have in mind by comparing the explanation patterns (R4) in manually created subpopulations (R2). 

To instantiate this pipeline, we developed {\systemname}, an interactive widget embedded in the Jupyter notebook. In this way, data scientists can interactively analyze local explanations using code and conventional interactions such as clicking and brushing. Moreover, the intermediate result of the visual analysis can be exported as a variable (e.g., DataFrame and list in Python) for further analysis. 

%% file: section/05-vis.tex
\section{\systemname: Visualization and Interaction}
\label{sec:vis}
\begin{figure*}
    \centering
    \includegraphics[width=\textwidth]{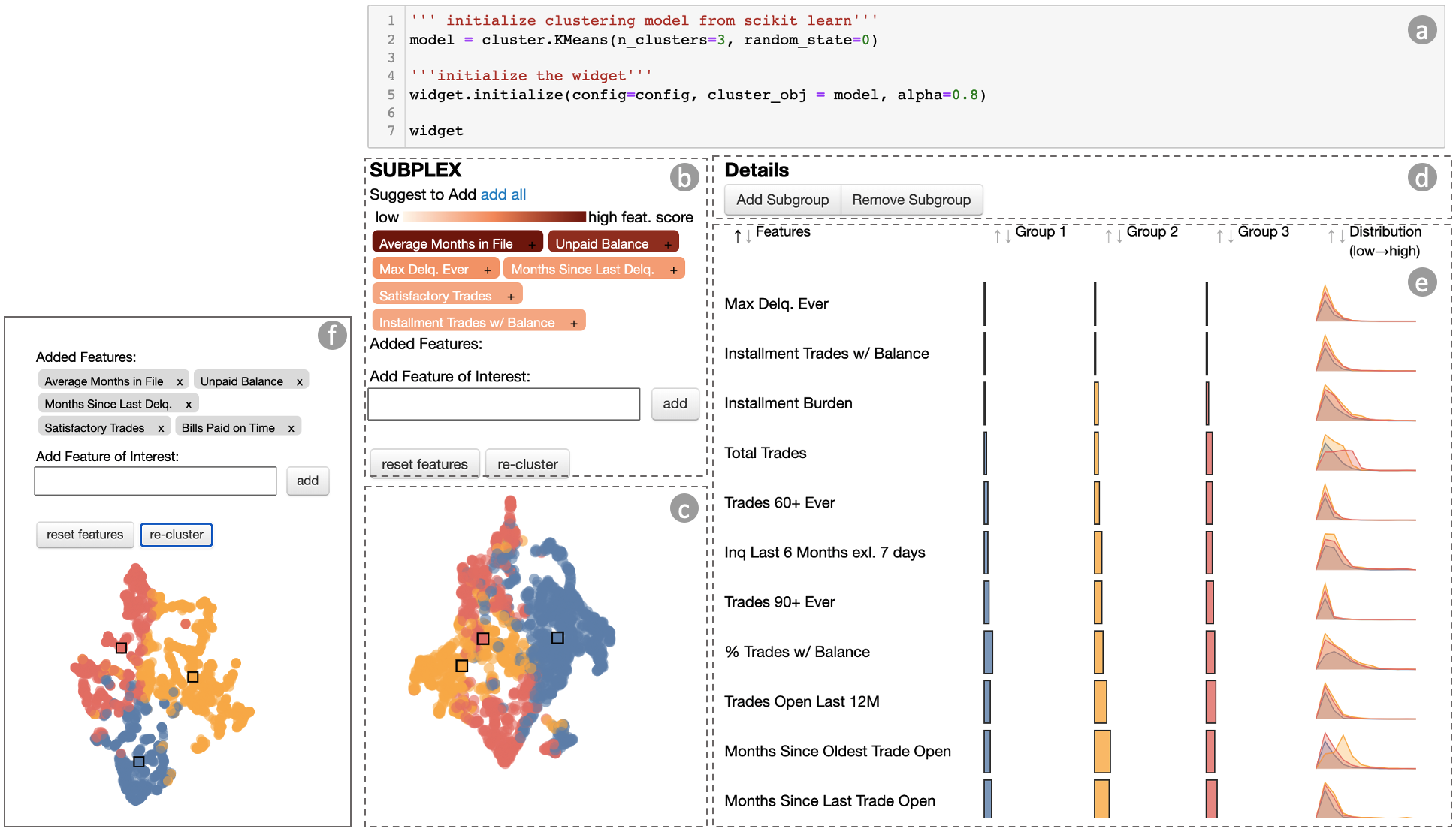}
    \caption{{\systemname} contains five linked views: (a) code block, (b) cluster refinement view, (c) projection view, (d) subpopulation creation panel, (e) local explanation detail view.}
    \label{fig:ui}
\end{figure*}

As shown in Figure~\ref{fig:ui}, the {\systemname} includes five views to assist users to understand and analyze local explanations: (a) the \textit{code block} inside a Jupyter notebook enables users to set user-defined parameters and export analysis results as variables; (b) the \textit{cluster refinement view} provides feature suggestions and allows users to refine clusters; (c) the \textit{projection view} shows the distribution of local explanations; (d) the \textit{subpopulation creation panel} enables manual add-and-remove for subpopulations; (e) the \textit{local explanation detail view} displays the aggregated local explanation distribution for each subpopulation.

In the following subsections, we introduce the coordinated views and interactions according to how they support the three stages in the pipeline and fulfill the design requirements.

\subsection{Generation}
The widget is initialized with the feature names, an array of local explanations, and the automatic clustering model users prefer to use. As shown in Figure~\ref{fig:ui}a, the user inputs the feature names and local explanations in \texttt{config}, feed in the widget a pre-defined \texttt{cluster\_obj}, and specify \texttt{alpha} ($\alpha$) for later cluster refinement. Then the widget renders below the code block \revise{(R5)}.

In the \textit{cluster refinement view} (Figure~\ref{fig:ui}b), we display the top 6 features with the highest feature scores \revise{(R1). Users can directly add the recommended features by clicking ``+'' or typing the feature names to refine clusters.}

\begin{figure*}
    \centering
    \includegraphics[width=.9\textwidth]{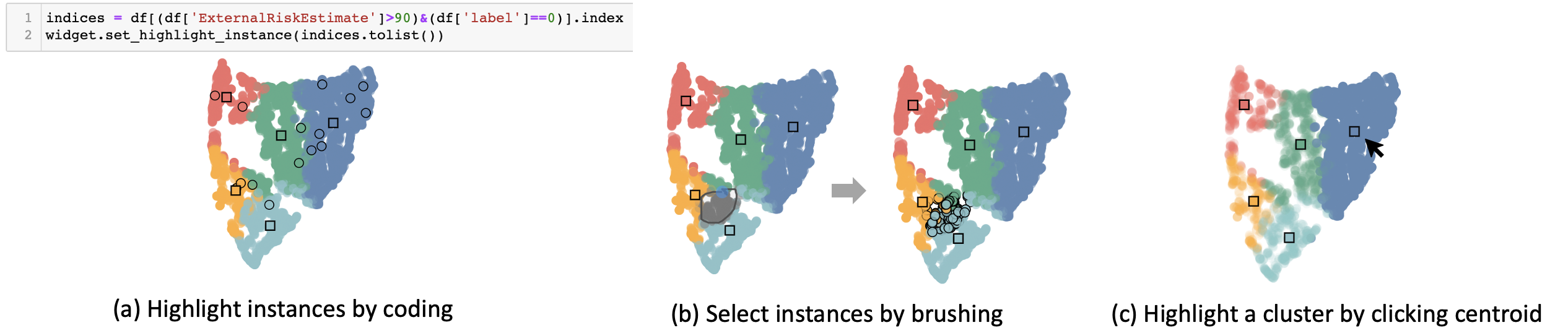}
    \caption{Three methods of selecting/creating a subpopulation for inspection.}
    \label{fig:selection}
\end{figure*}

\subsection{Exploration}
The \textit{projection view} (Figure~\ref{fig:ui}c) presents the local explanation distribution in a 2D space. Based on the weighted local explanations, we generate a 2D position for each local explanation vector using UMAP~\cite{mcinnes2018umap}. With the distribution of local explanations, users can gain a general idea of how the clusters distribute and how large each cluster is \revise{(R3)}. 

In the \textit{local explanation detail view} (Figure~\ref{fig:ui}e), each row shows the explanations for one feature. The first column is always the feature name, which follows the bars representing the average local explanation values in different subpopulations. The last column of each row displays the histogram of the local explanation values in different subpopulations. We enable feature sorting based on explanation values in one ore more subpopulations to facilitate pattern identifications \revise{(R3)}. Users can click the ``Distribution'' column to sort the features based on the local explanation variances among the clusters. \change{In both the \textit{projection view} and the \textit{local explanation detail view}, we use the same categorical color scheme to encode different subpopulations.
}


As shown in Figure~\ref{fig:selection}, users can define subpopulations in three ways \revise{(R2)}. The first method is specifying the instances of interest. By running \texttt{set\_highlight\_instance()} in the \textit{code block}, the user-defined instances will be highlighted in the projection. The second method is brushing instances in the \textit{projection view}. The third method is clicking the centroid points (squares in the projection) to highlight all the instances in the selected cluster.
These selection methods allow users to quickly acquire their desired set of explanations for analysis. \revise{By default, any selected dots are highlighted with a black edge as shown in Figure~\ref{fig:selection}-a, b. However, if users focus on a specific cluster that usually contains a lot of points, showing all the selected points with black edges will cause visual clutter. Thus, we fade the dots in unselected clusters to highlight the selected cluster as shown in Figure~\ref{fig:selection}c.}


\subsection{Interpretation}
We enable visual comparison of local explanation patterns in multiple subpopulations in the \textit{local explanation detail view}. In this view, the juxtaposition of bars enables visual comparison of the aggregated local explanations in different groups \revise{(R4)}.

\revise{To support the comparison with user-defined subpopulations (R2), the tool automatically calculates the average local explanation of a user-defined group.}
As shown in \revise{Figure~\ref{fig:loan_code}b}, the average value of the selected instances is shown as rectangles with black edges, while the average explanation value for an existing group/cluster is shown as a rectangle without black edges. After clicking the \textit{Add Subgroup} button (Figure~\ref{fig:ui}d), a new column of bars for this subgroup will be added. To confirm the hypotheses about the relationship between feature values and the local explanations, users can export the feature value histogram by calling \texttt{plot\_val\_histogram()} (Figure~\ref{fig:case_loan}b), and check the local explanations for the feature with specific values by calling \texttt{plot\_customized\_expl()} (Figure~\ref{fig:case_loan}c).

{\systemname} enables users to export the analysis results and store them as a variable in the Jupyter notebook \revise{(R5)} by compiling code in the \textit{code block} as shown in Figure~\ref{fig:loan_code}-c,d,e. 

\begin{figure}
    \centering
    \includegraphics[width=\columnwidth]{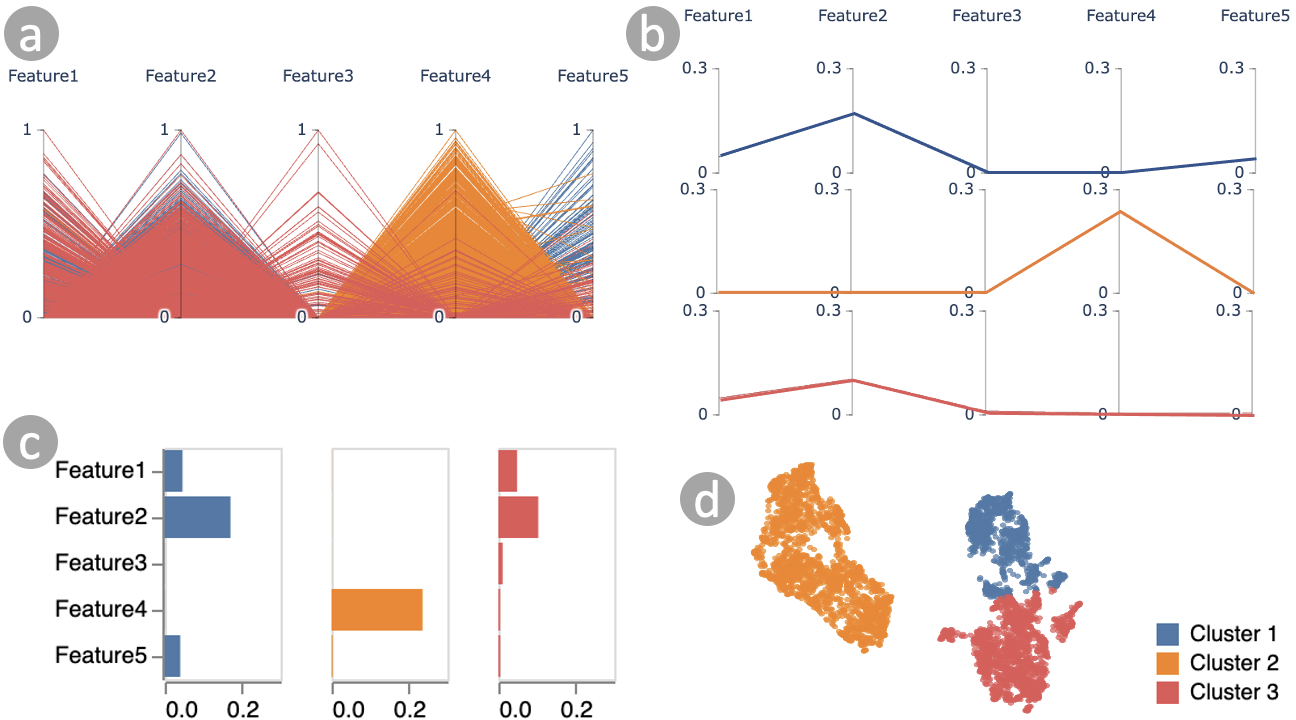}
    \caption{\change{Alternative designs for subpopulation analysis: (a) a parallel coordinates plot for all the explanations, (b) a parallel coordinates plot for aggregated explanations, (c) bar charts for aggregated explanations, (d) projection of explanations in a scatter plot.}}
    \label{fig:alternatives}
\end{figure}

\change{
Several design alternatives for the subpopulation analysis were considered (Figure~\ref{fig:alternatives}). 
The first one was to use a parallel coordinates plot (Figure~\ref{fig:alternatives}a).  Although it can show all the local explanations in a subpopulation, this plot easily becomes cluttered, which makes it hard to interpret the pattern in a subpopulation. To help with pattern identification (R3), we can visualize aggregated local explanations (e.g., mean values) as shown in Figure~\ref{fig:alternatives}b. Compared with parallel coordinates, aligned bar charts (Figure~\ref{fig:alternatives}c) have stronger color stimulus and are conventional. Therefore, we use bar charts to visualize aggregated local attributions. Moreover, compared with these charts providing specific explanation values, a projection of local explanations (Figure~\ref{fig:alternatives}d) can better demonstrate the spatial relationship and distribution of multiple subpopulations. As a result, we include the \textit{projection view} in \systemname with various interactions to support visual analysis.
}

%% file: section/06-eval.tex
\section{Evaluation}
\label{sec:eval}
To demonstrate how {\systemname} facilitates local explanation analysis, we first present a usage scenario of analyzing local explanations for a loan application model. Then we demonstrate how the steerable clustering and projection techniques enable local explanation analysis for high-dimensional data such as text. In the end, we describe the feedback on {\systemname} from eight domain experts.

\begin{figure}
    \centering
    \includegraphics[width=.5\textwidth]{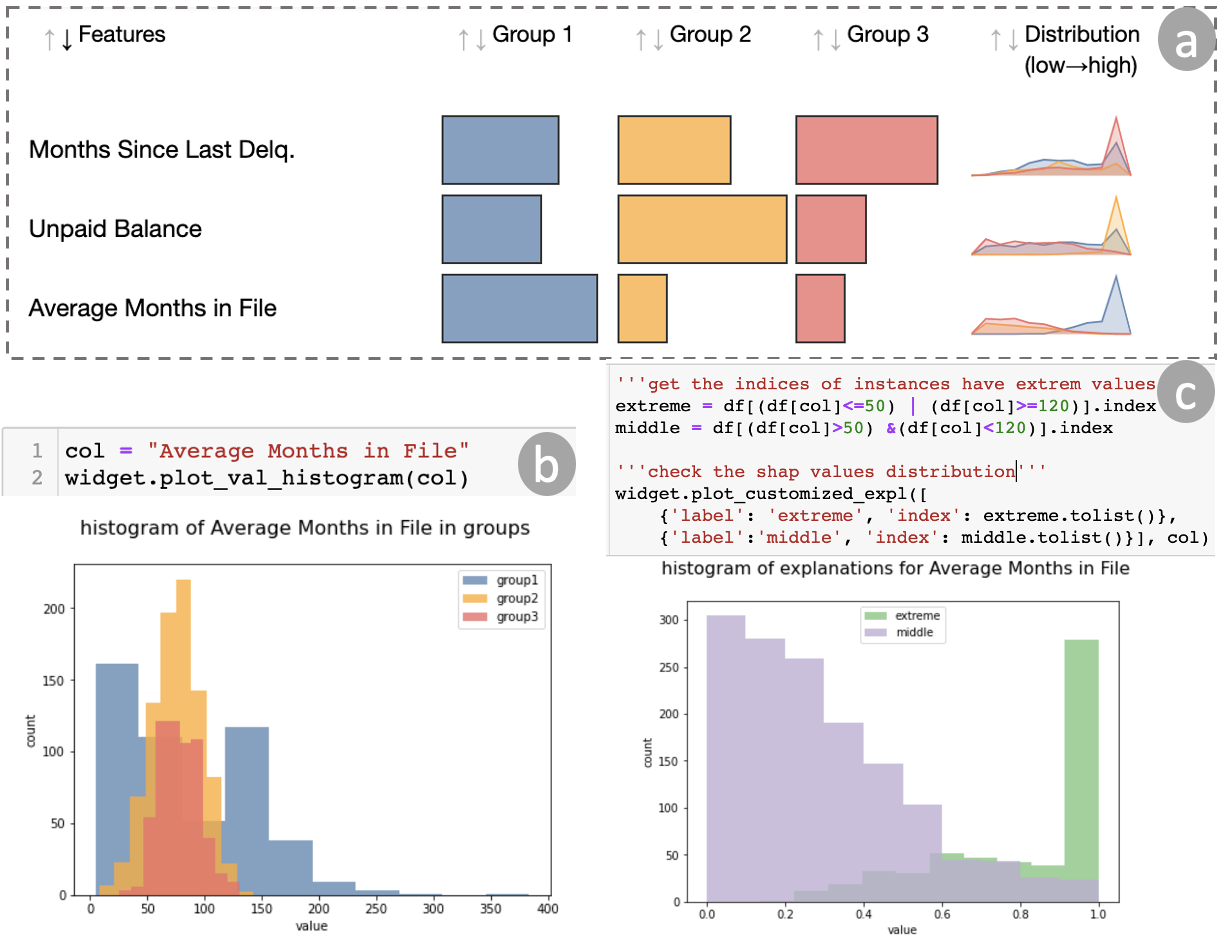}
    \caption{The analysis of local explanation patterns for a loan application dataset. }
    \label{fig:case_loan}
\end{figure}

\begin{figure*}
    \centering
    \includegraphics[width=.9\textwidth]{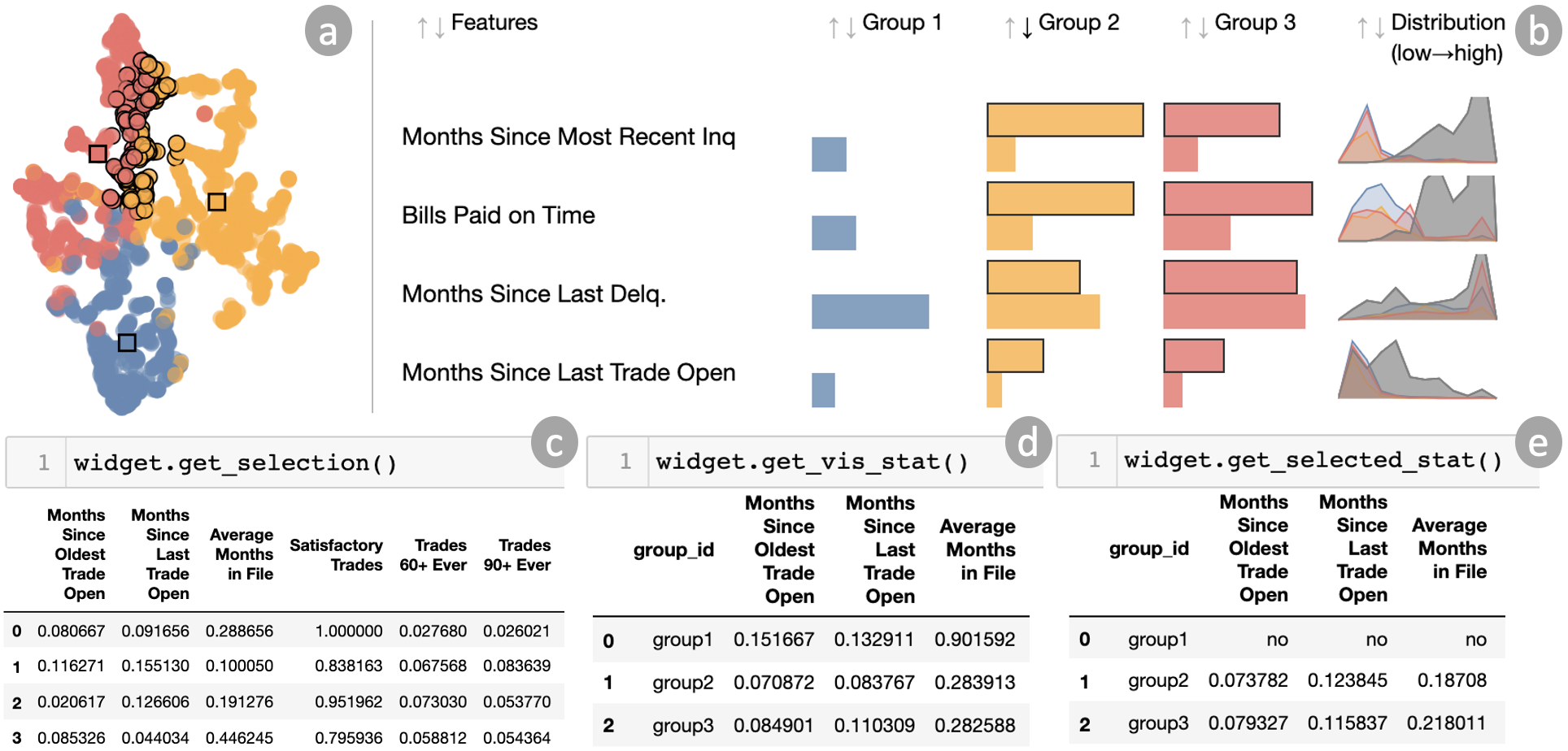}
    \caption{Brush selection and then output the aggregated information as a DataFrame in Python.}
    \label{fig:loan_code}
\end{figure*}

\subsection{Usage Scenario: Loan Application}

A model developer, Alice, trains an SVM model on a set of loan applications~\footnote{https://community.fico.com/s/explainable-machine-learning-challenge}. This model predicts whether a loan applicant will default or not with a test accuracy of $73.6\%$. Before the model deployment, Alice wants to understand how the model behaves. Thus, Alice generates SHAP values for all the test instances for analysis, which takes $0.37s$.

\textit{Cluster Generation and Refinement.} 
After running the widget, the initial results of the automatically generated clusters are not well separated in the projection view (Figure~\ref{fig:ui}c) due to many features (i.e., 22) in the explanation vectors. Thus she browses the feature suggestions (Figure~\ref{fig:ui}b) and adds three features \textit{Average Months in File}, \textit{Unpaid Balance}, and \textit{Months since Last Delq.} and manually selects \textit{Satisfactory Trades} and \textit{Bills Paid on Time} (R1). Then she re-generates the clusters and gets more separated results (Figure~\ref{fig:ui}f). 
\revise{Throughout the analysis, the widget takes $5s - 8s$  to generate clusters and calculate the projection for 1975 instances.}

\textit{Pattern Identification.}
By looking at the table (Figure~\ref{fig:case_loan}a), she finds some patterns (R3): \textit{Months Since Last Delq.} is important among all the three clusters. However, \textit{Unpaid Balance} has a strong influence in Group 2. Also, \textit{Average Months in File} has a strong influence in Group 1. Thus, Alice wants to understand why these features provide different ML models' rationale for the dataset.

\textit{Confirm Hypothesis.}
Alice first exports the histogram (R5) for the \textbf{actual values} of \textit{Average Months in File} in the three groups. For group 1, the values are between 0 to 400, with two peaks around 50 and 120 (blue bars in Figure~\ref{fig:case_loan}b). Yet, group 2 and group 3 have small values, mostly between 50 and 120. Alice now has a hypothesis that the feature \textit{Average Months in File} has a stronger influence when it has extreme values such as fewer than 50 or more than 120. To verify, Alice defines two groups of extreme values (either $<50$ or $>120$) and middle values (between 50 and 120) for \textit{Average Months in File} (R2, R5). Then she runs a query, and the widget outputs the histogram of the local explanation values for the two defined groups (R4, R5). As shown by the green bars in Figure~\ref{fig:case_loan}c, the extreme values of \textit{Average Months in File} have strong influences on the model predictions.



\textit{Pattern Identification.}
Alice wants to understand the local explanations around the group boundaries for further analysis. So she brushes the projection and selects boundary instances (Figure~\ref{fig:loan_code}a). In the Detail view (Figure~\ref{fig:loan_code}b), Alice observes how the local explanations of boundary instances are different from the discovered groups. It shows that the boundary instances are influenced much more by \textit{Months Since Most Recent Inquiry} and \textit{Bills Paid On Time} compared with other instances outside the boundary. 

\textit{Summary Export.}
After learning the model's behavior, Alice exports the summary visualization and the aggregated local explanation from different groups as a Python DataFrame (Figure~\ref{fig:loan_code}-c,d,e) for later discussion with her colleagues (R5). 

\subsection{Usage Scenario: Sentiment Analysis on Tweets}

A data scientist, Bob, is considering using an open-sourced model \textit{twitter-roberta-base-sentiment}~\cite{barbieri2020tweeteval} to predict whether a tweet contains a \textit{positive}, \textit{negative}, or \textit{neutral} sentiment. He runs the model on the test data~\cite{barbieri2020tweeteval} of 12284 tweets and gets an overall accuracy of $72\%$. He then wants to gain a general idea of how the model makes predictions. 

\textit{Cluster Generation and Refinement.} Bob first generates and stores the SHAP values~\cite{lundberg2017unified} of each instance for the class predicted by the model for analysis. It takes more than one hour to generate SHAP values for all instances with this large language model. For a text instance (e.g., a tweet), each token (word) is a feature. After an initial processing step of tokenization and lemmatization, the SHAP values for the whole test set can be transformed into an array of 12284 instances with 17000 features. After removing features (tokens) with low SHAP values and stop words (i.e., ``to'', ``in'', etc.), Bob starts the local explanation analysis based on 137 features.

\begin{figure}
    \centering
    \includegraphics[width=.5\textwidth]{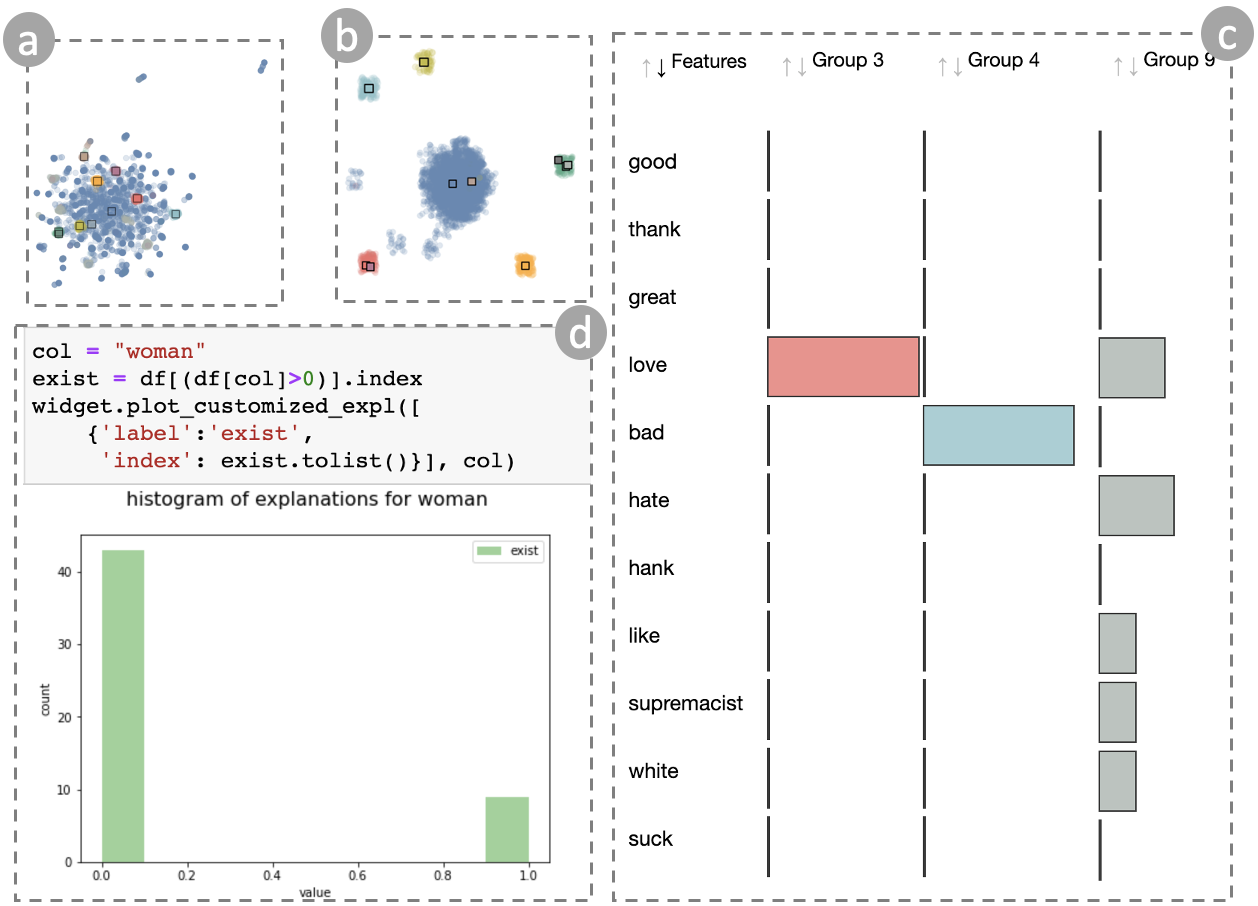}
    \caption{Usage scanario of understanding the sentiment analysis model of tweets.}
    \label{fig:case_nlp}
\end{figure}

Bob initializes {\systemname} and tries to generate 10 clusters at the beginning. Soon he realizes that the result does not contain a clear visual structure (Figure~\ref{fig:case_nlp}a). Thus, he checks the feature suggestions and notices that the tokens ``thank'', ``love'', ``good'', and ``bad'' have high feature scores while providing intrinsic meanings to him for understanding the sentiment in texts.
As a result, Bob selects these features and reruns the clustering (Figure~\ref{fig:case_nlp}b). The local explanations in the discovered clusters are well separated now.
\revise{For the inspected data (137 features/tokens, 12284 instances), it takes around $16.7s - 17.5s$ for the widget to generate clusters and calculate the projection.}



\textit{Pattern Identification and Interpretation.} Bob then inspects the local explanation detail view to extract the local explanation patterns. For example, Group 3 and Group 4 shown in Figure~\ref{fig:case_nlp}c, are mostly influenced by a single feature/token (e.g.,``love'', ``bad''); some groups are influenced by the combination of similar tokens (e.g., ``love'' and ``like''). 
As shown in Figure~\ref{fig:case_nlp}c, besides verbs related to emotions, tweets in Group 9 are influenced by non-emotional words, such as ``supremacist''. Bob then outputs the actual tweets and finds the tweets in this group are mostly complaints and labeled as negative sentiments. 

\textit{Hypothesis Testing.} During the inspection, Bob notices that the token ``woman'' has high SHAP values in some subgroups. So he has a hypothesis that the models learned a biased correlation between ``woman'' and the sentiment. To test this, Bob plots the SHAP value distribution of  ``woman'' and its variants (e.g., ``women'') for tweets that mention them by calling \texttt{plot\_customized\_expl()}. As shown in Figure~\ref{fig:case_nlp}d, there are around 50 tweets containing ``woman'' in total, where ``woman'' is extremely important in 10 tweets. Although the sample size is too small to conclude that the model learned a bias, more inspection is needed for cases related to gender bias in the language model. 


\revise{
\subsection{Evaluation via Expert Feedback}
We presented the widget to 8 domain experts in a commercial bank and collected their feedback. 
They are all data scientists specializing in financial data and used the Jupyter notebook in their work. They all had experience with model development, visualization, and local explanations. Among them, E2 is one of the machine learning experts who participated in the whole design process and helped us identify the design requirements, while the rest had no exposure to {\systemname} before. 

\textbf{Procedure.} Interviews were conducted remotely and lasted around 30 minutes each. We began each interview with an introduction, during which we clarified the goal of {\systemname} and provided a tutorial on the tool. Then we presented the case study of loan application as described in the first usage scenario to the experts. They were allowed to ask any questions or give any comments. In the final phase, we conducted an interview that incorporated several questions about the overall usefulness, comparison with the tools they were using, and general pros and cons of {\systemname}.

\textbf{Usefulness.} In general, the experts found {\systemname} useful. They liked the visualization of the automatic recommendations and the drill down into subpopulations of interest. The scatter plot and color encoding were intuitive for them to understand subpopulations. E1 stated that grouping the local explanations instead of aggregating the overall feature importance score would be more accurate. E2 mentioned that {\systemname} helped bridge the gap between global explanations (too high-level) and individual local explanations (too many details). E3 liked the cluster visualization and found the workflow of the subpopulation discovery useful. E6 particularly liked the visual design where everything is approachable and interactive within a notebook environment, making it easy to integrate into other work. Similarly, E8 liked the interactivity most because it allowed more exploratory analysis of model development with interpretability.

\textbf{Feature/Instance of Interest.} The experts had different preferences of features and subpopulations in exploring the local explanations. E1 expressed interest in the boundary instances. They wanted to understand how these instances ended up in different clusters. E3 preferred to explore age-related features to ensure they were not biased on age. Similarly, E6 would like to inspect sensitive demographic populations regarding age, race, gender, etc. E8, however, focused more on the features that are non-monotonic to the target.

\textbf{Comparison with other tools.} 
The tools or methods used by the experts were mostly static plots generated via code or exclusive web applications. Compared with other methods, all of the experts liked the interactivity of {\systemname}, which made the exploration more powerful. Moreover, E4 mentioned that compared with Accumulated Local Effects (ALE) Plot and Partial Dependence Plots (PDP), which focus on just one or two features in one chart, {\systemname} provided a more global view of data and explanations. E5 said that, in their typical work, they only scan through output automatically generated across all features/segments, while {\systemname} provided intelligent pre-selection in terms of suggesting features or clusters.

\textbf{Suggestions for improvement.}
To further improve {\systemname}, the experts suggested we provide more details on the dimensionality reduction method. E1 commented that different projection methods could change the analysis outcome. E5 wanted to include more details about the cluster/group creation details in the output because this can be easier to reproduce the results when discussing with other teams. We also found the comparison of local explanation methods a good next step in this work. For example, E2 wanted to explore and interpret the local explanation of \textit{Bills Paid on Time} because, in another project, this feature was one of the most important features to change for a counterfactual explanation.  

}

%% file: section/07-discuss.tex
\section{Discussion}
\label{sec:discuss}

\subsection{Design Guidelines}
Expert feedback on {\systemname} was positive. They liked the interactivity and found the visualization intuitive for them. We attribute these comments to our design process, where we took the workflow and analysis goal into consideration through the hierarchical task abstraction. We expect the presented hierarchical task abstraction can be helpful for future model explanation analysis tools.

Our target users are data scientists who have experience with static plots generated via coding. Therefore, we adapted the visualizations such as bar charts and scatterplots, which are mostly used by data scientists in their daily work. We also noticed that data scientists had some questions about the local explanations and the methods for clustering and projection. To better support the analysis, designs should be more transparent about the metrics and models used in the tool.

\subsection{Lessons Learned}
We discuss the lessons learned with respect to the following aspects:

\textbf{Interacting with data in a coding environment is powerful.} 
Our participants appreciated the interactive exploration of local explanations in Jupyter notebook. $87.5\% (7/8)$ of our participants preferred a widget like {\systemname} in the coding environment. Only one participant did not have a strong preference. We expect more visual analytics designs for data scientists embedded in the coding environment so that the users do not need to switch between different platforms.

\textbf{Importance of the human-in-the-loop pipeline with intelligent UI.}
We also confirm the importance of involving humans in the loop with the assistance of an intelligent UI for local explanation analysis. Participants liked the guided clustering with feature suggestions while enabling manual feature adding. Participants expressed quite different interests in the features and instances for inspection, which requires the UI to support flexible feature/instance selection. We expect to see more work integrating human and machine intelligence in local explanation analysis.


\subsection{Limitations}
The system currently has certain limitations. 
First, our tool was developed with domain experts from a financial company, which might not represent all use cases concerning local explanations. 

Second, the current widget needs to calculate clustering and projection, which requires some computation overhead. In our use cases, the computations took less than 10 seconds for around 2,000 rows/ 20 columns and less than 20 seconds for about 1,2000 rows/ 100 columns. We chose the clustering (K-Means) and projection (UMAP) techniques that are not quadratic in time complexity, but other similar methods like hierarchical clustering or tSNE might not scale well. 


\change{Finally, by only interviewing eight experts from the financial domain, we may be overgeneralizing our results. For the next step, we are interested in a controlled study that more directly compare subpopulation-level model explanation analysis with different visual representations, such as parallel coordinates and bar charts, as well as showing aggregated local explanations or individual local explanations in a subpopulation.}

%% file: section/08-limit+conclude.tex
\section{Conclusion}
\label{sec:conclude}
In this work, through an iterative design process with expert machine learning researchers and practitioners, we propose steerable clustering and projection techniques to identify and select important features for an ML model's interpretation. We also identified a list of tasks for explaining a machine learning model using local explanations, which led to the design of {\systemname}, a novel visual analytics tool in the Jupyter notebook environment.
We demonstrate the effectiveness of our approach with two use cases and collect feedback from domain experts.